\documentclass{spie}
\usepackage{graphics}
\usepackage{tikz}
\usetikzlibrary{shapes,arrows}
\usetikzlibrary{shapes,snakes}
\begin{document}
\title{Stellar Intensity Interferometry:\\ Imaging capabilities of air Cherenkov telescope arrays}

\author{Paul D. Nu\~nez\supit{a}\footnote{\hspace{0.5cm}pnunez@physics.utah.edu}, Stephan LeBohec\supit{a}, David Kieda\supit{a}, Richard Holmes\supit{b},\\ Hannes Jensen\supit{c}, Dainis Dravins\supit{c}
\skiplinehalf
\supit{a} University of Utah, Dept. of Physics \& Astronomy, 115 South 1400 East,\\  Salt Lake City, UT 84112-0830, USA\\
\supit{b} Boeing LTS, Inc., 535 Lipoa Parkway, Suite 200, Kihei, Hawaii 96753-6907, USA\\
\supit{c} Lund Observatory, Box 43, SE-22100 Lund, Sweden\\
}

%\author{Paul Nu\~nez\\Stephan LeBohec\\David Kieda\\\emph{University of Utah}\\Richard Holmes\\
%Dainis Dravins\\\emph{Lund Observatory}}
\maketitle

\begin{abstract}
Sub milli-arcsecond imaging in the visible band will provide a new
perspective in stellar astrophysics. Even though stellar intensity interferometry was
abandoned more than 40 years ago, it is capable of imaging and thus
accomplishing more than the measurement of stellar diameters as was previously
thought. Various phase retrieval techniques can be used to reconstruct actual
images provided a sufficient coverage of the interferometric plane is
available. Planned large arrays of Air Cherenkov telescopes will provide
thousands of simultaneously available baselines ranging from a few tens of
meters to over a kilometer, thus making imaging possible with unprecedented
angular resolution. Here we investigate the imaging capabilities of arrays
such as CTA or AGIS used as Stellar Intensity Interferometry receivers. The
study makes use of simulated data as could realistically be obtained from
these arrays. A Cauchy-Riemann based phase recovery allows the reconstruction
of images which can be compared to the pristine image for which the data were
simulated. This is first done for uniform disk stars with different radii and 
corresponding to various exposure times, and we find that the uncertainty in reconstructing 
radii is a few percent after a few hours of exposure time. Finally, more complex 
images are considered, showing that imaging at the sub-milli-arc-second scale is possible.
%The same is done for a range of exposures so as to evaluate the
%sensitivity performance for various degrees of image complexity.
\end{abstract}

\section{Introduction}

There has been a recent interest in the revival of Stellar Intensity Interferometry (SII)
due to the excellent baseline coverage of planned  air Cherenkov telescope arrays\cite{holder2, Dainis0}. This interest 
has lead to developments in instrumentation, experimentation, and simulations of the capabilities of 
this technique. Various analog and digital correlator technologies\cite{Dravins.timescale} are being implemented by LeBohec 
et. al.\cite{stephan.spie}, and cross correlation of streams of photons with nanosecond scale resolution 
has already been achieved. The suitability of various proposed array configurations is being evaluated by Jensen et al.\cite{hannes.spie} to understand their different sensitivities for interferometric imaging before final choices of the array layouts are made.
% Array optimization is currently being performed by Jensen et. al. \cite{hannes.spie}, and this will aid in array design by maximizing the available Fourier plane coverage among existing telescope configuration options. 
Image reconstruction algorithms such as
the one suggested by Holmes et. al.\cite{Holmes} have opened the possibility of imaging and have
become interesting subjects in their own right. In
this paper we will focus on the imaging capabilities and limitations of 
air Cherenkov telescope arrays used as high angular resolution intensity interferometers.\\

High angular resolution astronomy in the optical range will open a whole new
field of exploration. The possibility of viewing stars as extended objects will
enable the testing of many current astrophysical models, and the knowledge
acquired will have consequences on many fields  related to stellar
astrophysics. As a first example consider the measurement of stellar diameters,
 which can be performed to an accuracy of a few percent with
the methods discussed in this paper (see section \ref{disks}). When measuring 
diameters at different wavelengths, we can learn about the behaviour of the optical depth as a 
function of the radius of a particular star \cite{Mozurkewich}. This type of measurement
becomes particularly useful at shorter wavelengths ($\lambda\sim400nm$) than those feasible with 
conventional amplitude (Michelson) interferometry.
Another interesting science case is the study of fast rotating stars, for which we can measure
oblateness, pole brightening, and disk formation.  There is also the study of interacting binaries. An
actual image of  an interacting binary will not only aid in the determination
of the orbital parameters to a high degree of accuracy,
but also in the study and imaging of mass transfer and accretion. Although not
rigorously discussed in this paper, imaging mass transfer will improve our
understanding of late stellar evolution, i.e. the formation of type II
supernovae, which are of great importance in cosmology \cite{review}, and our understanding
of the formation of compact objects. The number of interesting science cases and
astrophysical targets is overwhelmingly large, and a detailed discussion is
given by  Dravins et al. \cite{Dainis}.\\

We propose the use of Intensity Interferometry for high angular resolution
astronomy (see Holder et. al.\cite{holder2} for
more details). This technique was introduced in the 1950's by R. Hanbury Brown
\cite{Brown, Brown2} and implemented in the 1960's with the Narrabri Stellar Intensity
Interferometer\cite{Brown3}, accomplishing the measurement of over thirty stellar
diameters. The use of planned air Cherenkov telescope arrays poses a unique
opportunity to revive Intensity Interferometry. With hundreds of telescopes
separated by up to $\sim 1km$, it will be possible to have an unprecedented
coverage of the Fourier plane and thus achieving sub-milliarcsecond
resolution.\\

The most important difference between SII and amplitude interferometry is that
SII relies on the correlation between the low frequency intensity fluctuations
and so does not rely on the relative phase of the individual waves received at
different telescopes. Intensity interferometry measures the squared modulus of
the complex degree of coherence $|\gamma|^2$.

\begin{equation}
|\gamma|^2=\frac{<\Delta I_i \,\Delta I_j>}{<I_i> <I_j>}
\end{equation}

Here $<I_i>$ is the time average of the intensity received at a particular
telescope $i$, and $\Delta I_i$ refers to the low frequency intensity
fluctuations received at telescope $i$. Intensity interferometry has several advantages and
disadvantages when compared to amplitude interferometry. The main advantages are that
it is insensitive to atmospheric turbulence and that it does
not require high optical precision\footnote{For a more detailed discussion on
  the advantages of Intensity Interferometry see Hanbury Brown\cite{Brown3}}. The complex
degree of coherence is proportional to the Fourier transform of the object in
the sky (Van Cittert-Zernike theorem), and since we measure the modulus squared of
$\gamma$, the main disadvantage is that the phase of the Fourier transform is lost in the measurement
process, posing difficulties in recovering actual images from magnitude
information only. In addition to imaging difficulties, measuring a second order effect 
also results in sensitivity issues\cite{holder2}, which can be dealt with by using large 
light collection areas and exposure times. As for the imaging limitations, 
several phase retrieval techniques exist, and we will implement a two dimensional 
version of the one dimensional approach introduced by Holmes \& Belen'kii \cite{Holmes}. It is important
to note that our results pertain to a single phase recovery algorithm\cite{Holmes}, and a 
comparison to other algorithms is currently being investigated\cite{Holmes.spie}. Once 
a sufficient coverage of the Fourier plane is available, and phase
recovery is performed, a study on imaging capabilities can be performed. Here
we will first concentrate on the study of the uncertainty when reconstructing
disk-like stars. Then a less exhaustive analysis on the capabilities is
performed for more complicated images such as oblate rotators, binary stars
and stars with bright \& dark features.\\

 The outline of the paper is the following:  First we 
will briefly discuss the phase retrieval technique. Then
we will discuss  the simulation of our data and
how it will be fitted to an analytic function so that the phase retrieval
method can be applied. Finally we will discuss the capabilities for imaging
disk-like stars, binary stars and more complicated  objects.

\section{Phase reconstruction}

The objective of phase reconstruction is to recover the phase of the Fourier transform of the image
from magnitude information only\cite{Holmes}. The resulting image is then reconstructed up to an 
arbitrary translation and reflection. It is simpler to first understand phase retrieval in one dimension and then generalize
 to two dimensions . One possible route towards phase retrieval starts by first approximating the 
continuous Fourier transform $I(x)$ by a discrete one ($I(m\Delta x)=\sum_j \mathcal{O}(j \Delta \theta) e^{ijmk_0 \Delta x \Delta \theta}$, where $\mathcal{O}(\theta)$ is the image in the sky and $k_0$ is the usual wave vector).\
 Then the discrete Fourier transform can be expressed as a magnitude times a phasor ($I(z)=R(z) e^{i\Phi(z)}$ where $z\equiv e^{imk_0 \Delta x \Delta \theta}$ is complex). The most important step is then to apply the
theory of analytic functions i.e. the Cauchy-Riemann equations\footnote{The C-R equations can be applied because $I(z)$ 
is a polynomial in $z$, where $z\equiv e^{i\phi}$.}. These relate the phase $\Phi$ and the log-magnitude $\ln R$ 
along the real or imaginary axes. One can show by using the Cauchy-Riemann equations, that the phase differences
along the radial direction in the complex plane\footnote{If $\xi$ is the real axis and $\psi$ is the imaginary axis, then a difference along the radial direction is $\Delta\xi+i\Delta\psi$.} are directly related to the differences in the logarithm of 
the magnitude (see Holmes \& Belen'kii \cite{Holmes} for more details), so that integrating the Cauchy-Riemann equations directly does not 
immediately solve for the phase. In other words, phase differences along the purely real or imaginary axes are not available directly from the data.\\

Since $z$, the independent variable of the Fourier transform ($z\equiv e^{imk_0 \Delta x \Delta \theta}$), has modulus equal to 1, 
the phase differences that we seek lie along the unit circle in the complex plane. Consequently, the 
procedure to find the phase consists in first assuming a plausible solution form, then taking 
differences in the radial direction of the complex plane, and finally fitting the data to the radial differences of
the assumed solution. A general form of the phase  
can be postulated by noting that the phase is a solution of the Laplace equation in the 
complex plane (applying the Laplacian operator on the phase and using the Cauchy-Riemman equations 
yields zero). Since the phase differences
are known along the radial direction in the complex plane we can take radial differences of the 
general solution and then fit the log-magnitude differences (available from the data) to
the radial differences of the general solution.\\

One can think of this one-dimensional reconstruction as a the phase estimation along a single slice in the Fourier plane. A 
generalization to two dimensions can be made by doing the same procedure for several slices. The direction of the slices 
is arbitrary, however for simplicity we reconstruct the phase along horizontal or vertical slices in the Fourier plane, 
and noting that one can relate all slices with a single orthogonal slice, i.e. once the phase at the origin is set to zero, 
the single orthogonal slice sets the initial values for the rest of the slices. The resulting reconstructed phase will be arbitrary
up to a constant and a linear term, which corresponds to a translation. It should be noted that the above solution approach gave reasonably good results.  However, it is not the only possible approach.  We have also investigated Gerchberg-Saxton phase retrieval, Generalized Expectation Maximization, and other variants of the Cauchy-Riemann approach\cite{Holmes.spie}.

\section{Procedure}

Having briefly discussed the phase reconstruction algorithm, our basic procedure for recovering images is the following: First we simulate realistic data as would be obtained from an air Cherenkov telescope array such as CTA or AGIS. Once simulated data are available, they are fitted to an analytic function so that the phase recovery algorithm can be applied in a more straightforward way. Finally, once the phase is recovered, the inverse Fourier transform will provide us with a reconstructed image. Some details concerning the simulations and data fitting approach now follow.

\subsection{Simulation of realistic data and array design used} 

The simulation of the the data that may be produced by an array of telescopes starts from a pristine image, generally  $2048\times2048$ pixels with an arbitrary dynamic range. The squared magnitude of the Fourier transform of this image is obtained by means of an FFT algorithm and it is normalized so its maximum at zero baseline is equal to one. With a wavelength $\lambda=400\,nm$, and the full scale of the image typically set to 10\,mas, this provides a value for the expected degree of coherence on a square grid with a pitch of $\sim 8.2\,m$ extending over a $\sim 16.8\,km\times 16.8\,km$ area. \\

The squared Fourier magnitude map is sampled by the set of pairs available in the simulated array. Simulations presented here have been obtained with an array of $N=97$ telescopes, each with a light collecting area of $100\,m^2$ and a quantum efficiency $\alpha=0.30$ resulting in an effective area of $30\,m^2$. The telescopes are distributed in the field according to an early design of the CTA array shown in figure \ref{array}. Such an array provides a coverage of the interferometric (u,v) plane with $N(N-1)/2=4656$ baselines many of which are redundant.  The baselines are shown in figure \ref{array}. The degree of coherence recorded by each baseline is obtained from a linear interpolation between the closest four points in the Fourier magnitude table.  \\

The data recorded by a real array would be affected by the diurnal motion of the observed star which affects the effective baselines by projection. The average correlation must then be recorded for each baseline at time intervals short enough for the baseline change to be negligible. In the first simulation study reported here we have decided to avoid this complication and simulate data that would result from the observation of fixed stars at zenith so the effective baselines used are those shown in figure \ref{array} without any further projective distortion. The implications of this simplification choice are a less uniform sampling of the (u,v) plane compensated by smaller error bars on the degree of coherence from each baseline record. These two effects essentially cancel each other as long as small scale features in the (u,v) plane are not central to the analysis. The benefits from the simplification is a reduction in the volume of data to handle (each simulation produces a single record for each pair of telescopes) and eliminates further arbitrary parameters (such as the site latitude, celestial declination, range of hour angles and time interval between recordings). \\

Once the degree of correlation within each baseline has been obtained, Gaussian noise is added.  The Gaussian nature of the noise was tested with detailed simulation of a pair of photo-multiplier tube signals corresponding to a random stream of photons. The time integrated product of the two traces was Gauss distributed. The magnitude $m_V$ of the star is used to compute a spectral density ($m^{-2}s^{-1}Hz^{-1}$) according to $n=5 \times 10^-5 \times 2.5^{-m_V}$. The standard deviation of the Gaussian noise added to the pair of telescopes $(i,j)$ is calculated as $\sigma=n   \sqrt{A_i \cdot A_j \cdot \Delta f \cdot \Delta t /2}$ where $A_i$ is the effective light collection area of the $i^{th}$ telescope, $\Delta f$ is the signal band-width and $\Delta t$ is the observation duration.  For simulations presented in this paper $\Delta f=200MHz$ which is a realistic choice when considerations on air Cherenkov telescope optics and electronics are taken into account (See Holder \& LeBohec \cite{holder2} and LeBohec et al. \cite{stephan.spie}).  \\

\begin{figure}
\begin{center}
\hspace{1.0cm}\rotatebox{-90}{\includegraphics[scale=0.5]{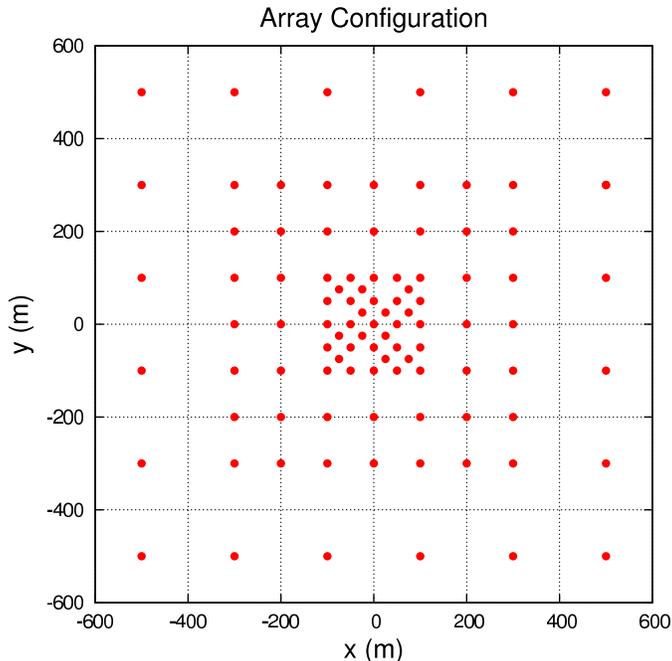}}
\vspace{0.5cm}
\caption{\label{array} Array configuration used for our analysis.}
\end{center}
\end{figure}

\begin{figure}
\begin{center}
\vspace{0.5cm}
\rotatebox{-90}{\includegraphics[scale=0.5]{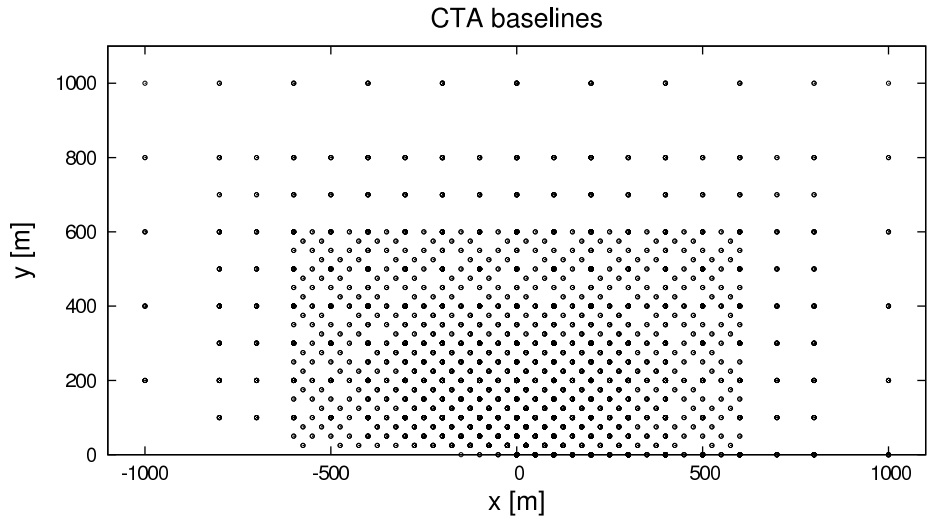}}
\end{center}
\vspace{0.5cm}
\caption{\label{array} Available baselines for the array design used in this study.}
\end{figure}

\subsection{Fitting the data to an analytic function}
The phase reconstruction algorithm is greatly simplified when data are known on
a square grid rather than in a `randomly' sampled way as is directly available
from observations. This is because sampling data on a fine square grid enables an
easier estimation of the derivatives of the log-magnitude.\\

Assuming that data $f(x_i)$ are known at positions $x_i$, with uncertainty
$\delta f(x_i)$, our goal is then to find a function that minimizes the
following $\chi^2$

\begin{equation}
\chi^2=\sum_i\left[\frac{(f(x_i)-\sum_ka_kg_k(x_i))}{\delta f(x_i)}\right]^2. \label{chi2}
\end{equation}

Here, the $a_k\,'s$ are the coefficients of the basis functions $g_k$ that we
want to use to fit our simulated data. Any complete basis will suffice in
theory, however it is more appropriate to choose a set of basis functions that
tend to zero at infinity. The reason for this requirement of our basis
functions is so that data are more realistically fitted in regions were there
is not much data available. For the case of the CTA array design that we used,
we noticed that there is less data at baselines greater than 600 m (see figure
\ref{array}). We found
that basis functions that meet this requirement are the solutions to the two
dimensional quantum-mechanical harmonic oscillator, i.e. Hermite polynomials
with Gaussian envelopes. These also turn out to be convenient because they are
eigenfunctions of the Fourier transform operator.\\

Now we can turn this problem into a linear system by taking derivatives with
respect to the unknowns $a_k$. Our data fitting typically starts with a small
number of basis elements, then we check to see if a certain reduced $\chi^2$
is met (in our case, we chose an acceptable reduced $\chi^2$ to be 1.5), if
this condition is not met, then the number of basis elements is increased
iteratively until it does.

\section{Imaging capabilities}

We may start by quantifying the resolution of an air Cherenkov telescope array such as the one
shown in figure \ref{array} by recalling that quantities that are related to each other
by a Fourier transform obey an uncertainty relation. For an order of magnitude
estimation it suffices to relate the size of the array to the maximum
resolution by

\begin{equation}
\Delta\theta\sim\frac{\lambda}{\Delta x}.
\end{equation}

With a kilometer size array ($\Delta x\sim 1\,km$), and a wavelength of $\lambda\sim400nm$ we obtain
a resolution of $\Delta\theta\sim0.1\,mas$. On the other hand, the largest
objects observable with an array whose inter-telescope separation is of the
order of $\Delta x\sim 50m$ is $\Delta\theta\sim1\,mas$. These order of
magnitude considerations will be taken into account when performing
simulations and image reconstructions, i.e. the minimum and maximum size of
pristine images will not go far beyond these limits.\\

We tested the imaging capabilities for simple objects, namely uniform
disk-like stars, oblate rotating stars, binaries, and more complex
images. First we will concentrate on the capabilities and
limitations for reconstructing uniform disk-like stars. We will show that such
a preliminary analysis reveals more precisely, when compared to the previous estimate, 
the sizes of objects that can be
observed. Even though using a Cauchy-Riemann based approach to recover images
might not be the most efficient way to measure stellar radii, such a study
will start to quantify the abilities of measuring other scale parameters in
more complicated images (oblateness, distance between binary components, etc.). 

%For each of these objects, one or more parameters sufficiently described the pristine image (for example: semi-major/minor axis, inclination angles, etc). We then investigated the similarity between real and reconstructed parameters for several images along with the uncertainty in their measurement.\\

\subsection{Uniform disks \label{disks}}

Simulated data were generated corresponding to uniform disk stars of a
particular brightness. We set the brightness to magnitude 6 after noting that
an error of a few percent in the simulated data can be achieved in a few
hours. Also, $6^{th}$ magnitude stars are appropriate since they roughly
correspond to the upper limit for most of the interesting astrophysical targets found by Dravins et
al. \cite{Dainis}. \\

\begin{figure}
  \begin{tikzpicture}
    \node (scatter){\rotatebox{270}{\scalebox{0.4}{\includegraphics{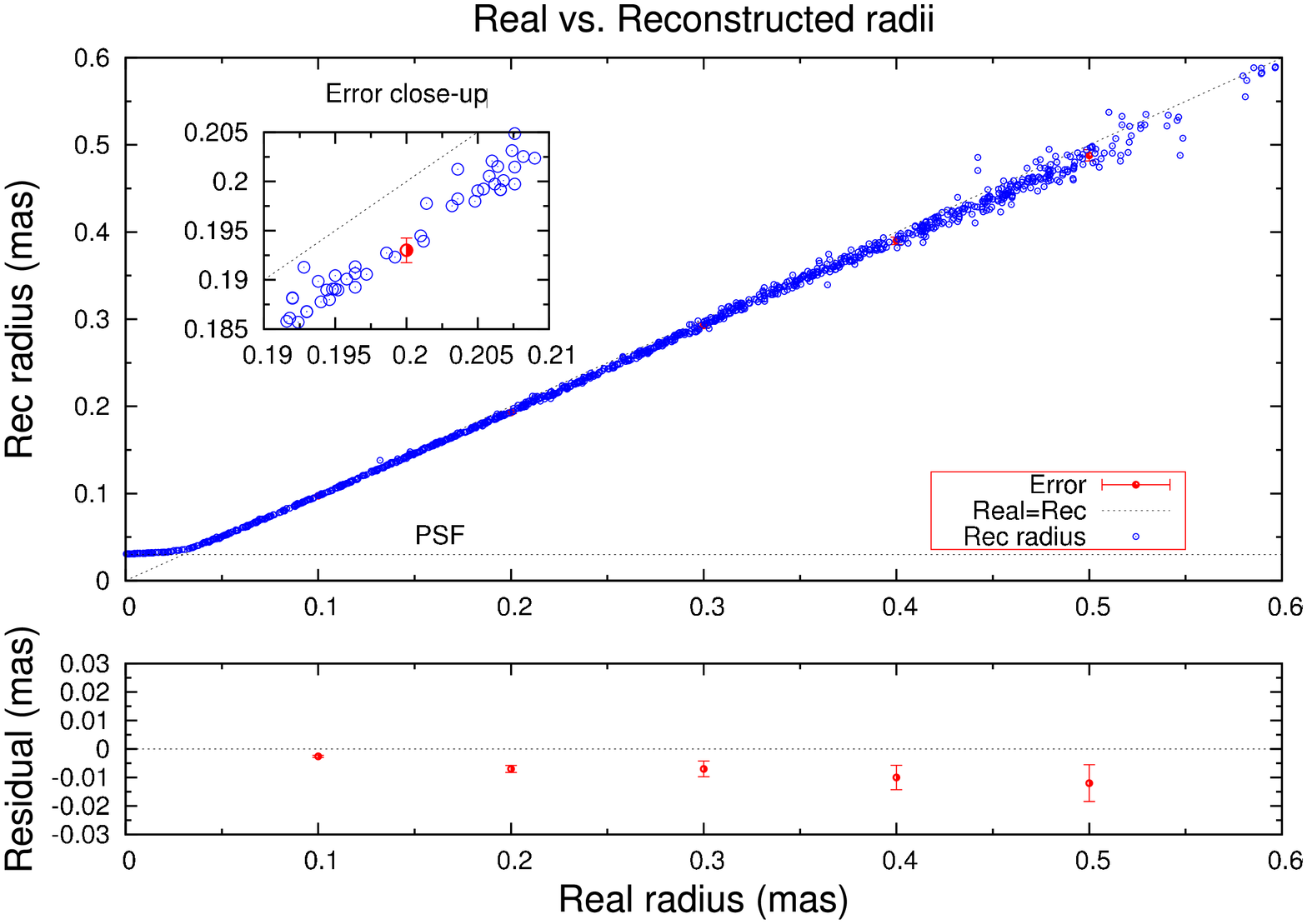}}}};
    \node [right of=scatter, node distance=8.5cm] {\rotatebox{-90}{\scalebox{0.4}{\includegraphics{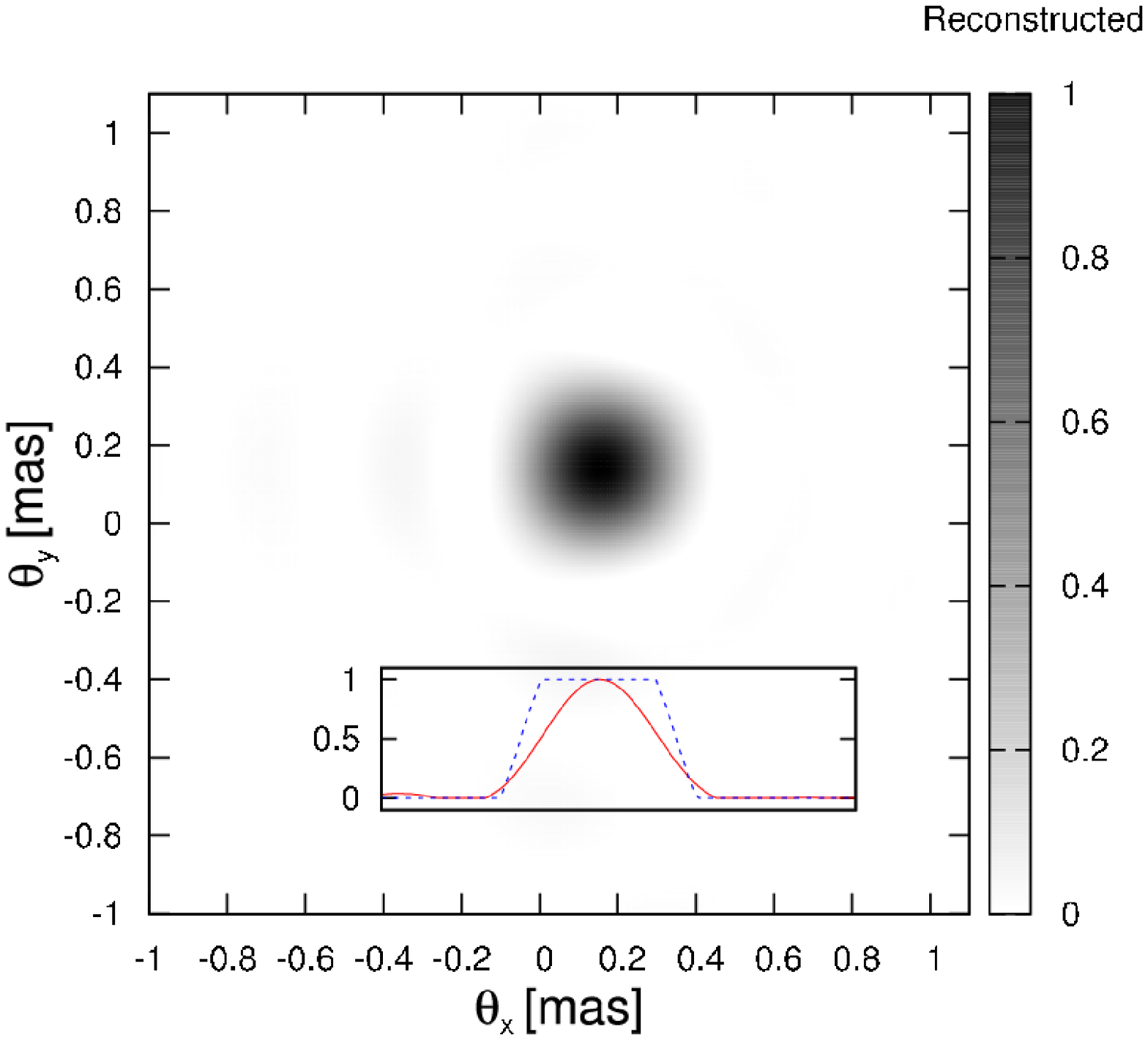}}}};
  \end{tikzpicture}
  \vspace{0.5cm}
  \caption{\label{radius} a) Simulated vs. Reconstructed radii for magnitude 6
    stars with 50 hours of observation time. The top sub-figure shows the
    uncertainty for a 0.2 mas measurement. The bottom sub-figure shows the residual (Reconstructed-Real) along with the uncertainty in the radius.
    b) Example of a reconstructed
    uniform disk of radius 0.2 mas. Also shown is a slice of the reconstructed
    image (solid line) compared to a slice of the pristine image (dashed line).}
\end{figure}

It is interesting to first study the uncertainty for a particular exposure
time and a brightness corresponding to $6^{th}$ magnitude.  We 
simulated data corresponding to uniform disks of random radii up to
$1\,mas$ for 50 hours of exposure time. An example of such a reconstruction is shown in figure
\ref{radius}b, where the brightness is shown in arbitrary units between 0 and
1. A first look at the example reconstruction reveals
that the edge of the reconstructed disk is not sharp, so a threshold in the
brightness was applied for for measuring the radius. The radius was measured
by counting pixels above a threshold and noting that the area of the disk is
proportional to the number of pixels passing the threshold. After experimenting
for different radii, we chose the threshold for measuring the radius to be
0.2. We can now compare the simulated and
reconstructed radii as shown in figure \ref{radius}a, where each point in the
figure corresponds to an individual simulation (including noise) and
reconstruction. \\

\begin{figure}
  \vspace{0.5cm}
$
\begin{array}{cc}
\rotatebox{-90}{\scalebox{0.35}{\includegraphics{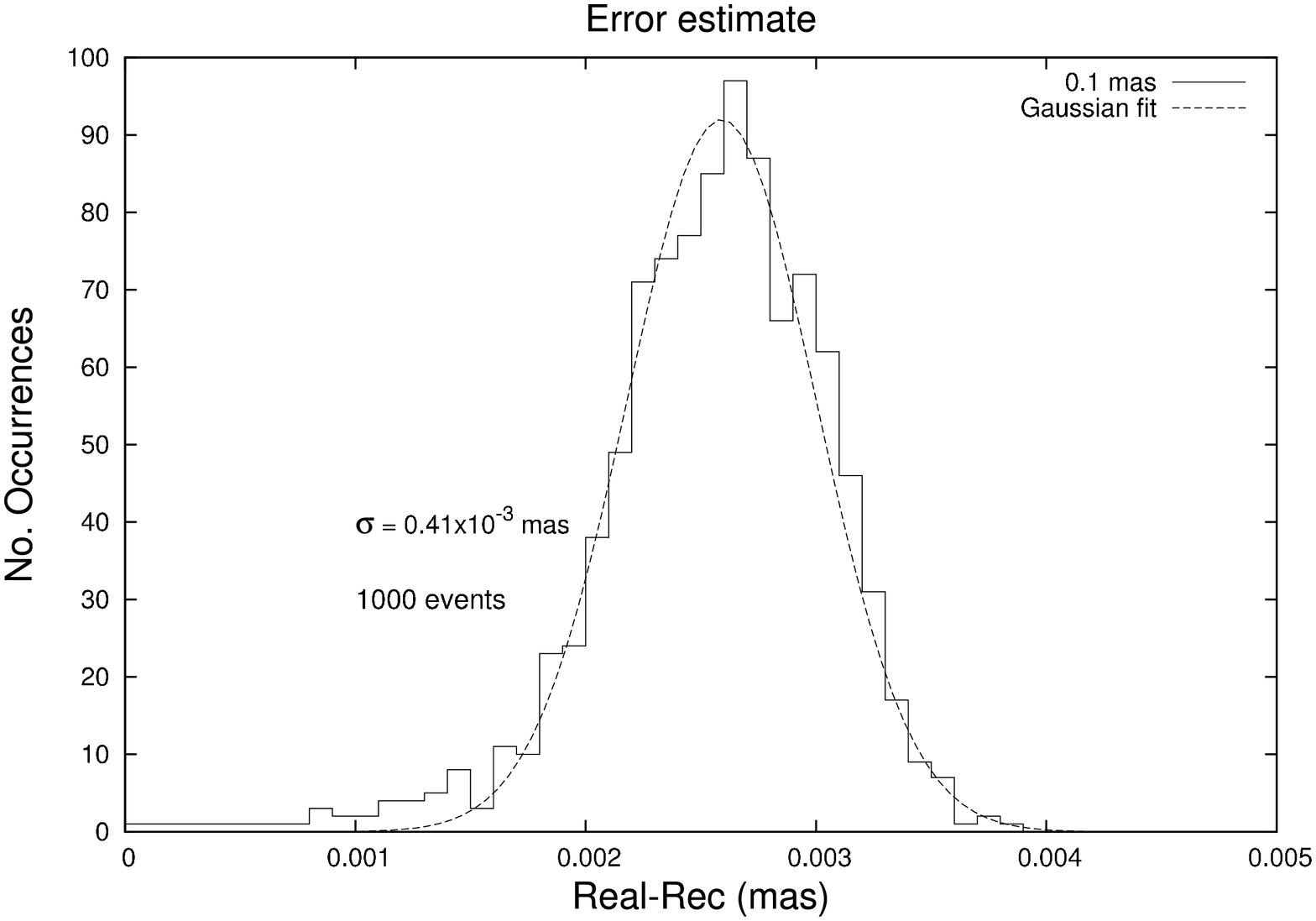}}}&
\rotatebox{-90}{\scalebox{0.32}{\includegraphics{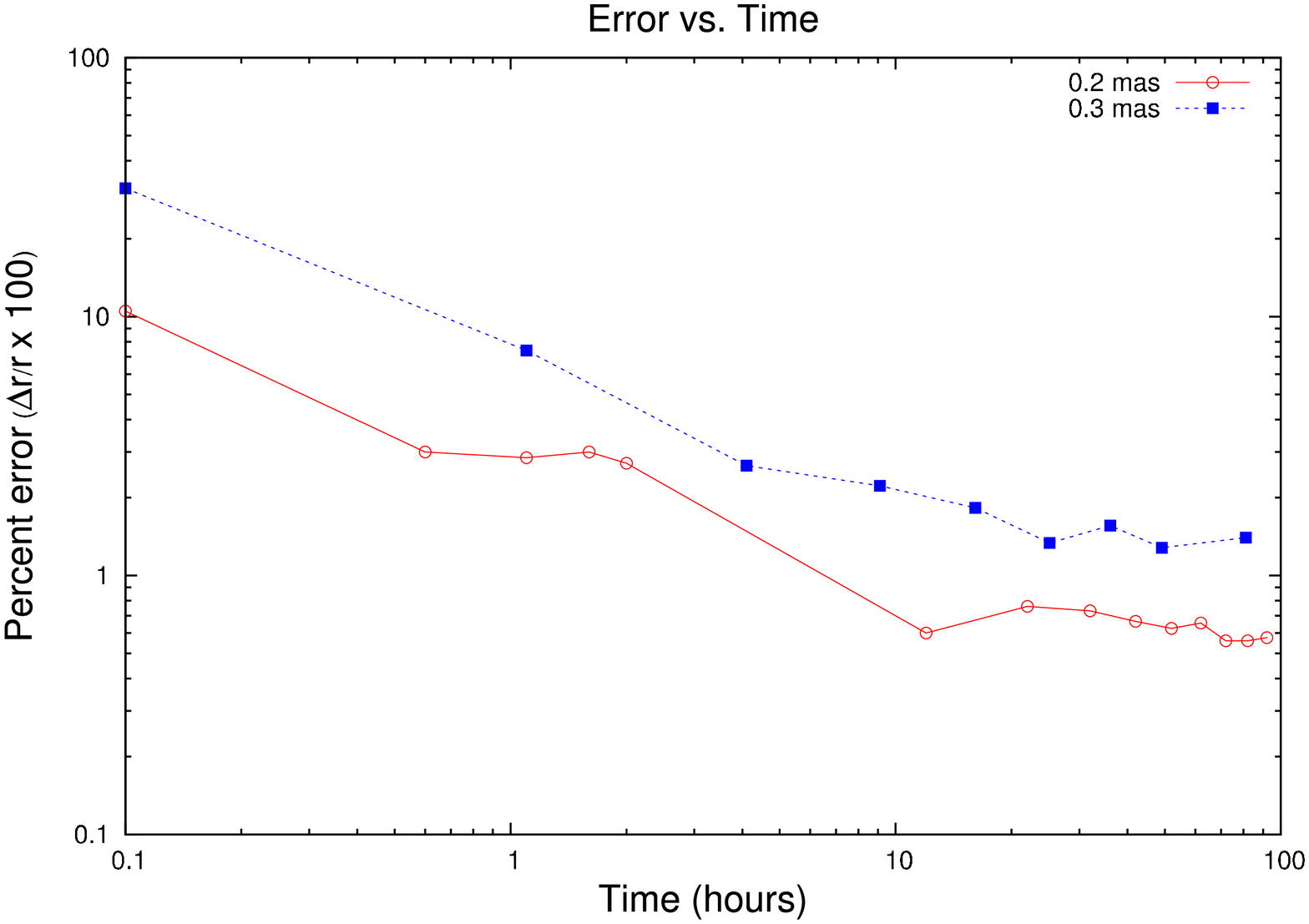}}} 
\end{array}
$
\vspace{0.5cm}
\caption{\label{uncertainty}a) Histogram of real radius minus reconstructed
  radius for 50 hours of exposure time on a $6^{th}$ magnitude star of $0.2\, mas$ radius. 
  b) Percent error as a function of time for several reconstructed radii.}
\end{figure}

Figure \ref{radius}a clearly shows that stellar diameters ranging from 0.05 mas
to 0.5 mas can be measured with uncertainties smaller than 5\%. The `tail' seen in the bottom left of figure \ref{radius}a shows the smallest measurable radius, so we take this radius ($\simeq 0.03\,mas$) to be the point spread function (PSF) of the array.  The uncertainty 
shown in the sub-figure in figure \ref{radius} was estimated after running noisy 
simulations many times as is shown in figure \ref{uncertainty}.\

It can be seen from figure \ref{radius}a, that the
uncertainty increases roughly linearly as a function of the pristine (simulated) radius. This
can be understood from the following argument: As the pristine radius
decreases, the distance to the first zero in the correlation increases as
$\sim r$, so the number of telescopes contained within the Airy
disk increases as $\sim r^2$. Consequently, decreasing the pristine radius is
equivalent to increasing the number of independent measurements by a factor of
$\sim r^2$ at most (at least by a factor of $r$). Since the
uncertainty will decrease as the square root of the number of independent
measurements, the error will decrease as $\sim r$ at most. For radii above
$0.6\,mas$, there are simply not enough baselines to constrain the Fourier
plane information for image reconstruction. For radii greater than $0.6\,mas$,
the size of the Airy disk is of the order of $100\,m$, and this results in
having less than 100 independent measurements as can be seen in figure
\ref{array}. Also shown in figure \ref{uncertainty} is the percent error as a
function of time for two different radii, where it can be seen that a percent error of less than
5\% is achieved after only a few hours.

\section{Complex images \label{complex}}

\begin{figure}$\begin{array}{cc}
\rotatebox{-90}{\includegraphics[scale=0.45]{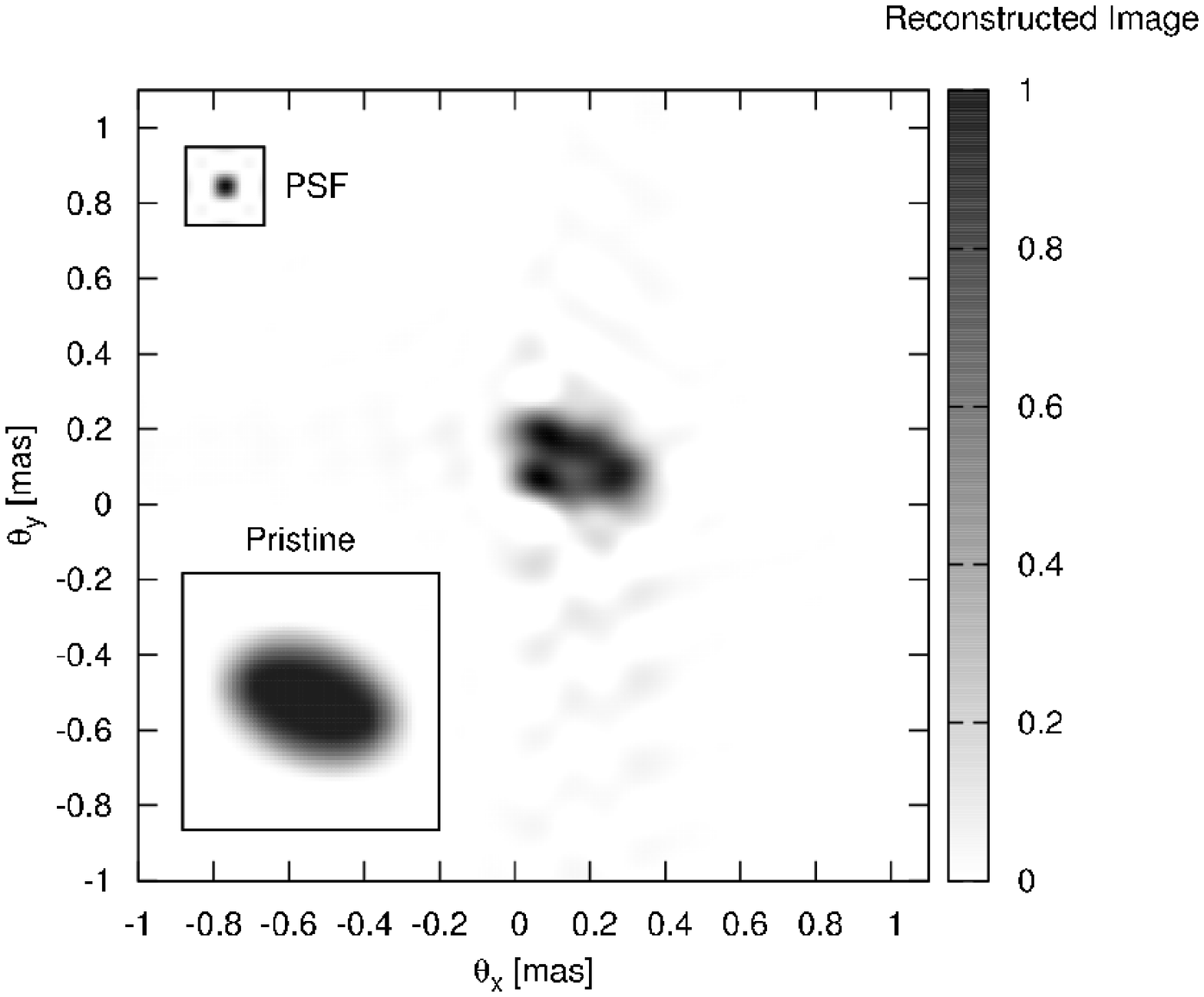}} &
\rotatebox{-90}{\includegraphics[scale=0.45]{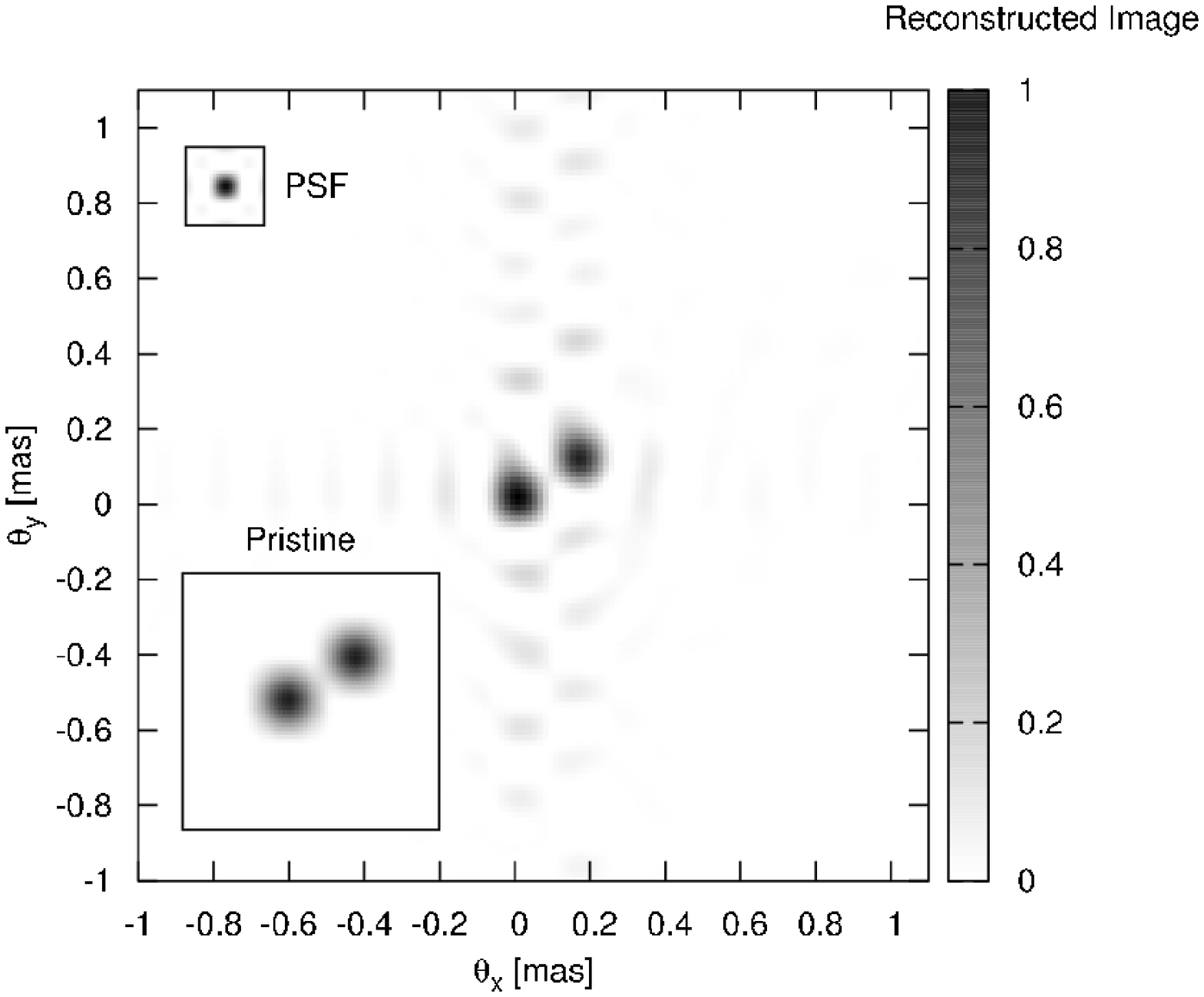}} 
\end{array}
$
\vspace{0.5cm}
\caption{\label{rotator and binary}a) Simulated and reconstructed oblate rotator of magnitude 3 and 10 hours
  of observation time. b) Simulated and reconstructed binary of magnitude 6
  and 12 hours of observation time.  }
\end{figure}

Our algorithm has also been tested on more complex images such as oblate rotating
stars, binary stars, and stars with brighter or darker regions. Since we have not developed a 
proper tool for quantitatively comparing simulated and reconstructed 
images\footnote{Since the simulated and reconstructed objects can differ by translations and reflections, developing a tool than can accurately quantify the difference between simulated and reconstructed images is not trivial.}, we will only show a few representative examples of what the algorithm is capable of.  Figure \ref{rotator and binary}a shows an example reconstruction of an oblate rotating star of
magnitude 3 and 10 hours of observation time, the semi-major axis and
semi-minor axis of the pristine ellipse are 0.2 mas and 0.12 mas
respectively. Also shown in this figure, and all subsequent ones, is the pristine 
image convolved with the point spread function of the array (see section \ref{disks}). Reconstructed 
noise of less than $10\%$ can be seen in the figure. These noise fluctuations are not so much a consequence of noise in the simulated data, but
of the reconstruction algorithm itself, and the preferential direction of this
reconstructed noise (along the vertical direction) is due to our choice of the
slice direction for the phase reconstruction, i.e. the phase was reconstructed
by taking horizontal one-dimensional slices of the magnitude, and then related
to each other with a single vertical slice. As for the structure (bumps) within the star in
figure \ref{rotator and binary}a, these start to appear when either the star
becomes bright or enough exposure time is supplied so that information other than the
first lobe in the Fourier magnitude is significant. In other words, when high
frequency portions are visible in the Fourier magnitude, fictitious structure
starts to become visible. This is most likely due to the fact that most of the high frequency 
information in the Fourier plane is used to reconstruct a dark background of several mas's with
a central bright region.\\% This leads us to believe that a better way of reconstructing high frequencies (structures within stars) is necessary. \\

The case of a binary star is shown in figure \ref{rotator and binary}b, corresponding
to a magnitude 6 binary star for 12 hours of exposure
time. The noise in the reconstructed image has the same origin as in the case
of the oblate star. Although the inclination angle was well reconstructed in
both cases in figure \ref{rotator and binary}, image reconstruction is
degraded when the symmetry axis is neither the x or y axis. Again, this is due
to the particular phase recovery method of taking horizontal or vertical
slices and the degradation is significantly reduced when aligning one
symmetry axis of the magnitude to our x or y axis.\\

\begin{figure}$\begin{array}{cc}
\rotatebox{-90}{\includegraphics[scale=0.45]{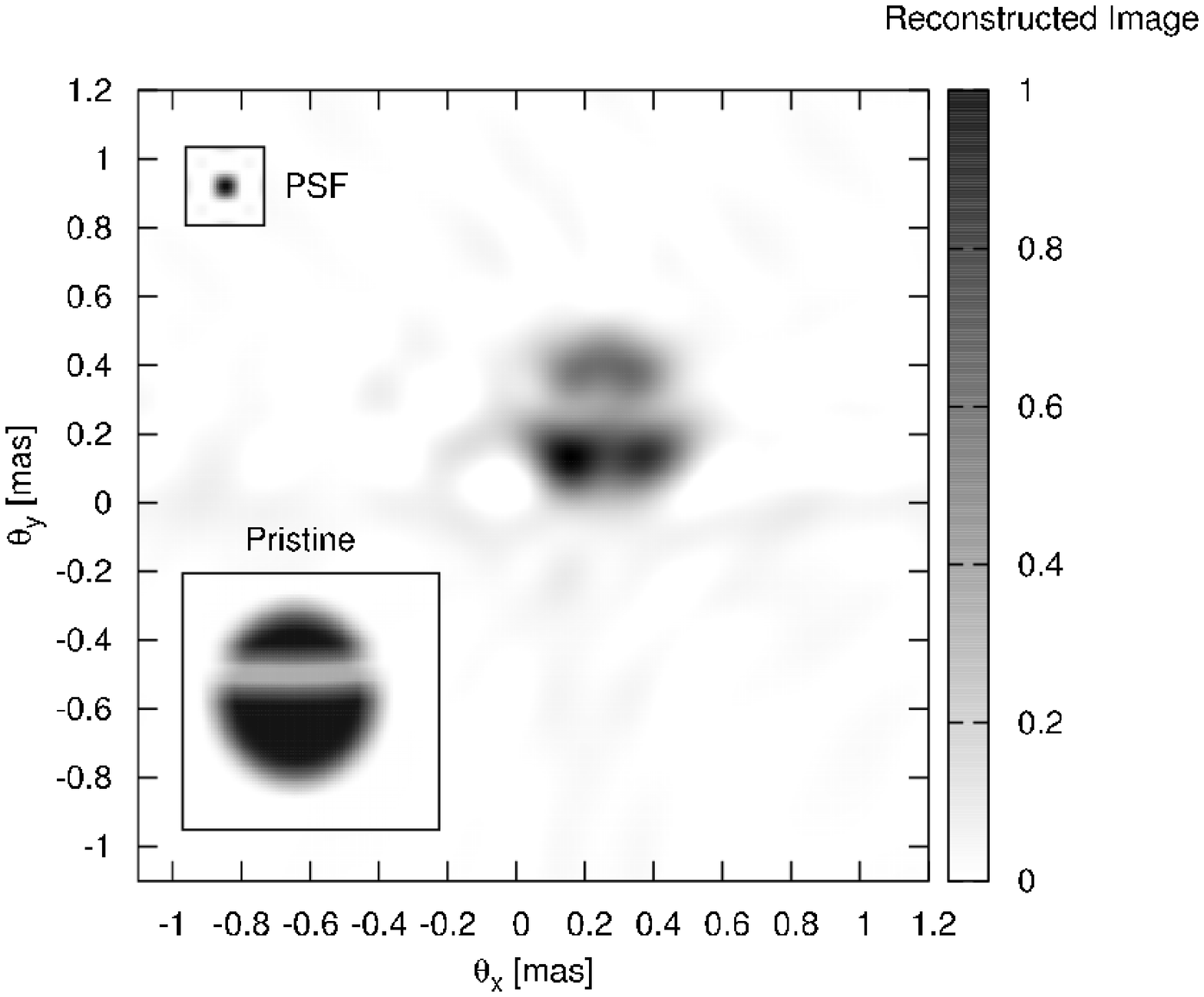}} &
\rotatebox{-90}{\includegraphics[scale=0.45]{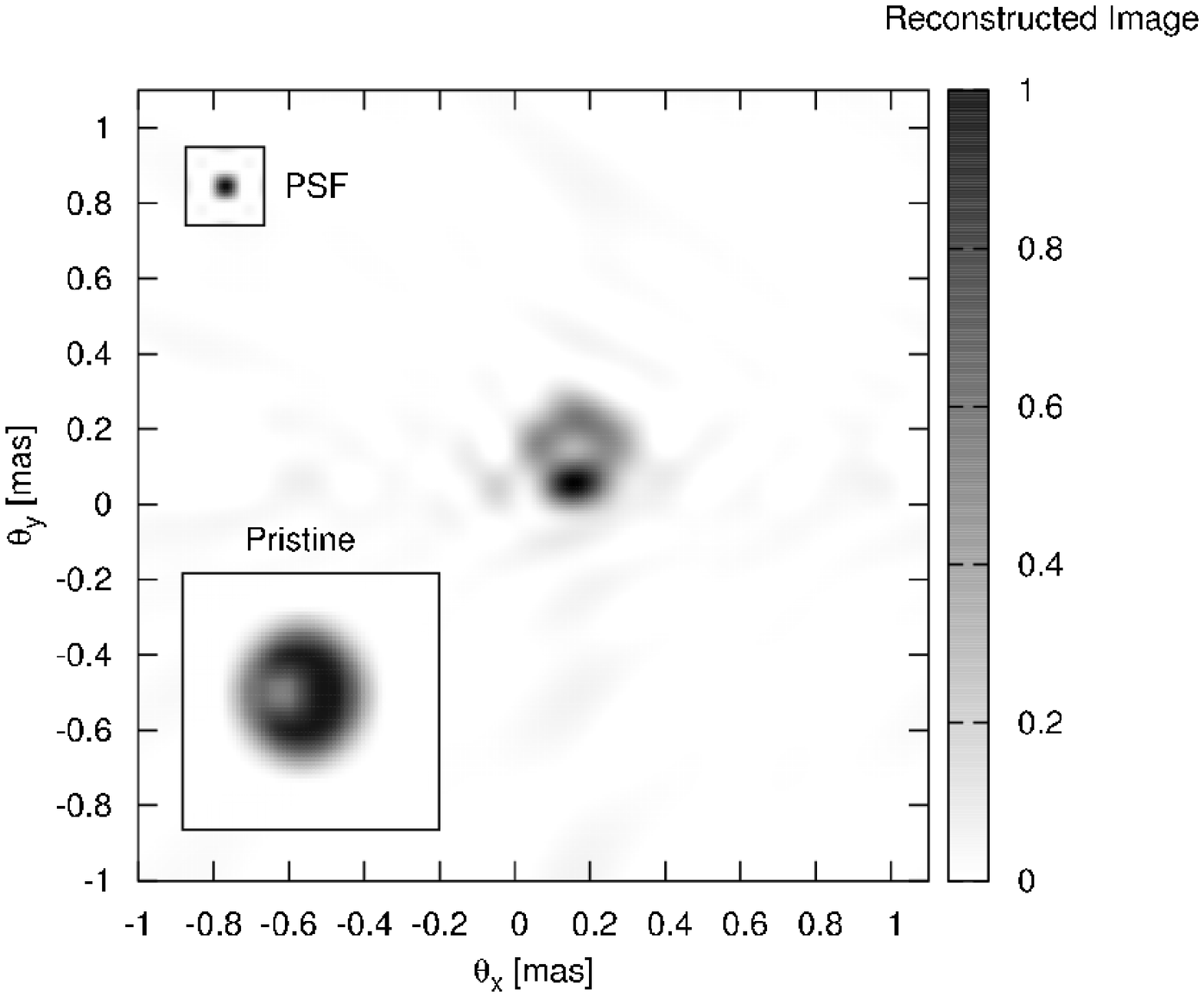}}
\end{array}
$
\vspace{0.5cm} 
\caption{\label{disk and spot} a) Simulated and reconstructed star obscured by a
  disk. This corresponds to magnitude 4 and 15 hours of observation time. b)
  Uniform disk of radius 0.2 mas with a dark spot of radius 0.05 mas. The
  simulation was done for magnitude 6 and 100 hours of observation time. }
\end{figure}

Having this symmetry consideration in mind, a slightly more complicated example
is shown in figure \ref{disk and spot}a, corresponding to a star obscured by a
disk (of dust for example). A black streak in the pristine image representing the obscuring disk is aligned 
with the $x$ axis. The black streak can be easily seen in the
reconstruction as well as the contour of the obscured star. This image becomes increasingly easier to reconstruct as
the image becomes more and more symmetric, that is, as the black streak in the pristine image moves
towards the center of the star\footnote{This can be understood by noting that the Fourier magnitude
  of a non symmetric binary becomes almost indistinguishable from the one
  corresponding to a central bright star with two fainter companions at either side.}. As a final example,
we considered the case of a dark spot in a star as shown in figure \ref{disk and spot}b. The simulation corresponds to 
a magnitude 6 star and 100 hours of observation. Even though 
the reconstructed image appears brighter in the bottom, and the location of the spot is more symmetric,
the size of the spot (comparable to the PSF) is well reconstructed.

\section{Conclusion and outlook}

We have shown that planned air Cherenkov telescope arrays have sufficient
baselines to provide an excellent coverage of the (u,v) plane. By first simulating
realistic data, we show that it is possible to achieve a signal to noise
ratio of the order of 10 within a few hours for relatively faint stars
(magnitude 6), and obviously higher S/N for brighter stars. Once data were simulated
we also show that imaging is possible using a Cauchy-Riemann \cite{Holmes}
phase recovery technique. A study of the error propagation with disk-like
stars reveals that the uncertainty in reconstructed radii is of a few
percent. We also explored imaging capabilities for more complex images
such as oblate rotating stars, binary stars, and stars with dark/bright
regions yielding good results. A quantitative analysis of the reconstruction
capabilities for complex images is still in progress. \\

The array design could certainly be improved by including telescopes at
shorter distances. This could significantly improve the size range of
observable objects, in particular, we could observe objects
of more than 1 mas across at 400 nm.\\

The analysis can also be improved by doing several things: If the pristine object has
a symmetry axis, a first reconstruction can be made to find it, and then a
second image reconstruction can be made to improve the first
results. Something else worth implementing is a first reconstruction only
constraining the low frequency components of the phase by using the low
frequency part of the magnitude, then a second reconstruction could be
performed, only dealing with the internal details of the image.\\

To conclude this simulation phase, pristine images generated from
astrophysical models should be generated in order to identify how much of the
astrophysical model can actually be constrained. This aspect of the simulation
phase is currently under development.\\

\textbf{Acknowledgements:}

This work is supported by grants SGER \#0808636 from the National Science Foundation. The work at Lund Observatory is supported by the Swedish Research Council and The Royal Physiographic Society in Lund.


\begin{thebibliography}{100}

\bibitem{holder2} Holder, J., and LeBohec, S. ``Optical intensity interferometry with atmospheric Cerenkov telescope arrays'' ApJ 649, 399-405 (2006)
\bibitem{Dainis0} Dravins D, LeBohec S, ``Towards a diffraction-limited square-kilometer optical telescope: Digital revival of intensity interferometry'' SPIE Proc. 6986, 698609 (2008)
\bibitem {Dravins.timescale}Dravins D., Hagerbo H. O., Lindegren L. et al., ``Optical astronomy on milli-, micro-, and nanosecond timescales'', Proc SPIE 2198, 289-301 (1994) % Dravins, D., et. al., 1994,  Proc SPIE 2198, 289.
\bibitem{stephan.spie} LeBohec, S., Adams, B., Bond, I., et al., ``Stellar intensity interferometry: Experimental steps toward long-baseline observations'', Proc. SPIE 7734, 7734-48 (2010)
\bibitem{hannes.spie} Jensen, H., Dravins, D., LeBohec, S. and Nu\~nez, P.D., ``Stellar intensity interferometry: Optimizing air Cherenkov telescope array layouts'', Proc.
SPIE 7734, 7734-64 (2010)
\bibitem {Holmes}  Holmes R. B. and  Belen'kii M. S. ``Investigation of the Cauchy–Riemann equations for one-dimensional image recovery in intensity interferometry'', J.Opt.Soc.Am A 21, 697-706 (2004)
\bibitem {Mozurkewich} Mozurkewich, D., Armstrong, J. T., Hindsley, R. B. et al., `` Angular diameters of stars from the Mark III optical interferometer'', AJ 126, 2502-2520 (2003)
\bibitem {review} van Dyk, S. D, ``Extragalactic binaries as progenitors of core-collapse supernovae'', New Astron. Rev. 48, 749-753 (2004)
\bibitem {Dainis} Dravins, D., Jensen, H., LeBohec, S. and Nu\~nez, P. D., ``Stellar Intensity Interferometry: Astrophysical targets for sub-milliarcsecond imaging'', Proc. SPIE 7734, 7734-9 (2010)
\bibitem {Brown} Hanbury Brown, R. and Twiss, R. Q., ``Interferometry of the intensity fluctuations in light I. Basic theory: The correlation between photons in coherent beams of radiation'', Proc. Roy. Soc. London A 242, 300-324 (1957)
\bibitem{Brown2} Hanbury Brown, R. and Twiss R. Q., ``Interferometry of the intensity fluctuations in light II. An experimental test of the theory for partially coherent light'', Proc. Roy. Soc. London A 243, 291-319 (1958)
\bibitem {Brown3} Hanbury Brown, R., [The Intensity Interferometer], Taylor \& Francis, London (1974)
\bibitem {Holmes.spie} Holmes, R., Nu\~nez, P. D. and LeBohec, S., ``Two-dimensional image recovery in intensity interferometry using the Cauchy-Riemann relations'', Proc. SPIE 7818B, 7818B-23, (2010)


\end{thebibliography}
\end{document}